\documentclass[prb,aps,twocolumn,showpacs]{revtex4-2}
\usepackage{amsfonts,amsmath,bm,multirow}
\usepackage[OT4]{fontenc}

\usepackage{graphicx}
\usepackage{xcolor}
\usepackage{epstopdf}
\usepackage{dsfont}
\usepackage{physics}
\usepackage{hyperref}

\newcommand{\kp}{\bm{k}\!\vdot\!\bm{p}}
\newcommand{\rr}{\bm{r}}
\newcommand{\kk}{\bm{k}}

\definecolor{bred}{HTML}{e31a1c}
\definecolor{bgreen}{HTML}{33a02c}
\definecolor{bblue}{HTML}{1f78b4}

\definecolor{armygreen}{rgb}{0.29, 0.33, 0.13}

\begin{document}

	\title{Phonon-assisted relaxation between triplet and singlet states in a self-assembled double quantum dot}
	\author{Krzysztof Gawarecki}
	\email{Krzysztof.Gawarecki@pwr.edu.pl}
	\author{Pawe{\l} Machnikowski}
	\affiliation{Department of Theoretical Physics, Wroc\l aw University of Science and Technology, Wybrze\.ze Wyspia\'nskiego 27, 50-370 Wroc\l aw, Poland}

	\begin{abstract}
		We study theoretically phonon-induced spin dynamics of two electrons confined in a self-assembled double quantum dot.  We calculate the transition rates and time evolution of occupations for the spin-triplet and spin-singlet states. We characterize the relative importance of various relaxation channels as a function of the electric and magnetic fields. The simulations are based on a model combining the eight-band $\kp$ method and configuration-interaction approach.  We show that the electron g-factor mismatch between the Zeeman doublets localized on different dots opens a relatively fast triplet-singlet relaxation channel. 
		We also demonstrate, that the relaxation near the triplet-singlet anticrossing is slowed down up to several orders of magnitude due to vanishing of some relaxation channels.
	\end{abstract}
	
	\maketitle
	
	\section{Introduction}
	\label{sec:intr}
	
	Isolated spins in solid-state systems offer fast optical control methods, long-time stability
	and high-fidelity conversion to photonic flying qubits  \cite{kroutvar04,Awschalom2018,Delley2017}. One of the possible implementations of solid-state spin qubits are single electrons in quantum dots (QDs),  which are optically active systems that offer the possibility of quantum coherent spin initialization, storage, readout, as well as entangling spins with photons \cite{Heiss2007,Press_NP10,muller12b,DeGreve2012,Schaibley2013,Gao2012}.
	A system composed of two coupled QDs (a quantum dot molecule, QDM) 
	offers an additional degree of freedom related to carrier localization, which can be controlled by an external electric field \cite{Krenner2005,Krenner2006}. In such systems, two electrons can be trapped in the ground-state manifold of the two dots, which allows one to encode the qubit in the singlet-triplet (S-T) space of spin states. For such a qubit, when operating at the S-T anticrossing, the dephasing effect of magnetic and electric field fluctuations is suppressed, leading to inhomogeneous coherence times exceeding 200~ns \cite{Weiss2012}. The singlet and triplet states in a QDM offer both spin-selective, as well as non-selective optical couplings to four-particle configurations, which are essential for state preparation and readout \cite{Elzerman2011,Delley2017,Delley2015a}.
	
	Both information storage and quantum state readout via light scattering rely on the stability of the spin configurations in the two-electron system.
	The relaxation rates between the singlet and triplet states were previously studied in gate-defined QDMs \cite{shen07} with the Dresselhaus spin-orbit (SO) coupling, where the spin relaxation times on the order of hundreds of micro-seconds were predicted for relatively low barriers and large dots, growing by orders of magnitude when the barrier height increases or the QD size decreases. In Ref.~\cite{Mutter2020}, it has been shown, that a difference between the site-dependent g-tensors in a QDM couples singlet to triplet states causing a leakage current.
	
	In self-assembled QDs, spin relaxation can be induced by many mechanisms resulting from band mixing and strain, out of which the shear-strain-induced spin-orbit coupling was shown to dominate the single-electron spin relaxation within the ground-state Zeeman doublet of a single QD \cite{Mielnik-Pyszczorski2018a}.
	The dependence of the strain, band-mixing, and spin-orbit effects on the QD geometry and composition profile requires precise modeling of carrier states and carrier-phonon couplings. Therefore, the methods and results relevant to gate-defined QDs are not directly transferable to self-assembled systems. 
	Reliable and computationally cost-effective modeling of self-assembled systems is possible using $\kp$ methods within the envelope function formalism~\cite{Ehrhardt2014}. This approach has been used to calculate spin-conserving relaxation between two-electron singlet states in self-assembled QDMs, yielding relaxation times on the order of tens of picoseconds~\cite{Gawarecki2010}. Single-electron spin relaxation between Zeeman sub-levels, modeled using the eight-band $\kp$ method, takes place on the time scales of $\sim 100$~ms at 1~Tesla, scaling as $B^{5}$ at low and moderate magnetic fields~\cite{Mielnik-Pyszczorski2018a}.
	
	In this work, we model theoretically phonon-induced relaxation between two-electron states in a self-assembled QDM. We calculate single-particle states using the $8$-band $\kp$ model with strain distribution found within continuous elasticity approach. The Coulomb coupling between the electrons is taken into account via configuration interaction (CI) method. Then the phonon-assisted transition rates are calculated using Fermi golden rule. We investigate the relaxation processes at various electric and magnetic fields, which can be used to control the system. We show that the difference between the $g$-factors in the two QDs, which naturally emerges due to their strain and geometrical characteristics, enhances the transitions from one of the triplet states, providing the dominant relaxation channel for a wide range of magnetic field. Thus, singlet--triplet spin relaxation in self-assembled structures is dominated by a real-space spin-orbit effect, which is relevant to transitions between states with the same $z$ component of the spin but belonging to different representations of the rotation group and therefore affects only many-particle states. The relaxation is strongly suppressed at the electric field corresponding to the minimum singlet-triplet splitting (the ``sweet spot"). We discuss also the kinetics of transition between the two-electron spin states and show that, depending on the magnetic field and temperature, various direct and sequential processes may contribute to the relaxation towards the singlet ground state.
	
	
	The paper is organized as follows. In Sec.~\ref{sec:model}, we describe the models used to calculate the single and double electron states, as well as the phonon-assisted relaxation. In Sec.~\ref{sec:results} we present the results of numerical simulations. Finally, Sec.~\ref{sec:concl} contains concluding remarks.
	
	\section{Model}
	\label{sec:model}
\begin{figure}[tb]
		\begin{center}
			\includegraphics[width=70mm]{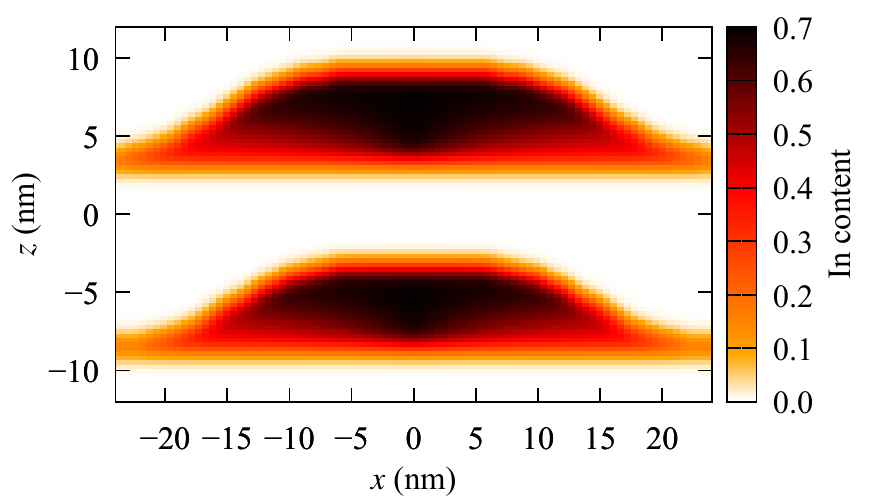}
		\end{center}
		\caption{\label{fig:comp}(Color online) In$_x$Ga$_{1-x}$As distribution in the QDM system. 
		}
	\end{figure}

	We consider two self-assembled, vertically stacked InAs/GaAs QDs~\cite{Krenner2005}. The composition gradient in the dots is modeled as a trumpet shape~\cite{jovanov11} with the maximum In content of 0.7 and the minimum of 0.4 (see Fig.\ref{fig:comp}). The mathematical models of the dot geometry and composition are described in Ref.~\cite{Karwat2019}. The heights of the lower (``l'') and the upper (``u") dot are taken $h_{\mathrm{l}} = 8.5 \, a_G$ and $h_{\mathrm{u}} = 9.5 \, a_G$ respectively, where $a_G = 0.565325$~nm is the GaAs lattice constant. The dots are placed on wetting layers of 0.4 In content and thickness of a single $a_G$. The parameters defining approximate QD base radii are chosen as $r_{\mathrm{l}}=28 \, a_G$ and $r_{\mathrm{u}}=30 \, a_G$. The material intermixing is accounted for via a Gaussian blur (where we took $0.6$~nm of the standard deviation). The simulations are performed for the axially oriented magnetic ($B$) and electric ($F$) fields.
	
	We calculate single-electron states $\ket{\phi_n}$ using the eight-band $\kp$ model~\cite{bahder90,Winkler2003}. We take into account strain distribution~\cite{pryor98b} and piezoelectric field~\cite{bester06a,Caro2015} in the system. The model and the details of its implementation are described in Ref.~\cite{Gawarecki2018a}.
	Finally, we calculate the two-electron states 
	$$
		\ket{\Psi_n} = \sum_{ij} c^{(n)}_{ij} \ket{\phi_i} \ket{\phi_j},
	$$
	where the coefficients $c^{(n)}_{ij}$ are found from exact diagonalization of the Coulomb interaction Hamiltonian within the CI approach. We also assume $c^{(n)}_{ij}=0$ for $j>i$. The CI basis contains 4 lowest single-electron states (i.e. two spin-dependent $s$-type states in each dot). In an idealized case (no spin-orbit interaction and no tunnel coupling), one can obtain the well known singlet/triplet configurations
	\begin{align*}
	\ket{S(0,2)} &= a^\dagger_{\mathrm{u} \uparrow} a^\dagger_{\mathrm{u} \downarrow} \ket{\mathrm{vac.}},\\
	\ket{T_+(1,1)} &= a^\dagger_{\mathrm{u} \uparrow} a^\dagger_{\mathrm{l} \uparrow} \ket{\mathrm{vac.}},\\
	\ket{T_0(1,1)} &= \frac{1}{\sqrt{2}} \qty[ a^\dagger_{\mathrm{u} \uparrow} a^\dagger_{\mathrm{l} \downarrow}  - a^\dagger_{\mathrm{l} \uparrow} a^\dagger_{\mathrm{u} \downarrow}  ]\ket{\mathrm{vac.}},\\
	\ket{T_-(1,1)} &= a^\dagger_{\mathrm{u} \downarrow} a^\dagger_{\mathrm{l} \downarrow} \ket{\mathrm{vac.}},\\
	\ket{S(1,1)} &= \frac{1}{\sqrt{2}} \qty[ a^\dagger_{\mathrm{u} \uparrow} a^\dagger_{\mathrm{l} \downarrow}  + a^\dagger_{\mathrm{l} \uparrow} a^\dagger_{\mathrm{u} \downarrow}  ]\ket{\mathrm{vac.}},\\
	\ket{S(2,0)} &= a^\dagger_{\mathrm{l} \uparrow} a^\dagger_{\mathrm{l} \downarrow} \ket{\mathrm{vac.}},
	\end{align*}	
	where $(N_\mathrm{l},N_\mathrm{u})$ describes the nominal occupations in the dots,  $a^\dagger_{\mathrm{(l/u)} (\uparrow / \downarrow) }$ is the creation operator for the electron state in the lower/upper dot and $\uparrow / \downarrow\,$ denotes the spin orientation.  
	Because of the SO coupling, 
	spin is no longer a good quantum number. Furthermore, spatial configurations are mixed by the tunnel coupling. However, spin-mixing effects are generally small, and the above notation is useful to classify well localized states (far from the tunnel resonances).
	
	The spin-related properties of many-particle states can be characterized by the $D_{j}$ irreducible representations of the full rotational group, where $j$ is related to the total angular momentum~\cite{Bir1974,Dresselhaus2010}. In the case of a two electron system, the direct product of representations gives $D_{1/2} \otimes D_{1/2} = D_{0} + D_{1}$. In consequence, states can belong to the one-dimesional trivial representation $D_{0}$ (singlet states) or to the three-dimensional $D_{1}$ (threefold degenerated triplet states). In the presence of the SO coupling, the geometrical symmetry breaking affects also the spin degree of freedom. For the QDM considered here, at $B=0$~T the system is described by the $C_{2v}$ symmetry point group. In this group, $D_{1}$ splits into one- and two-dimensional representations~\cite{Bir1974,Dresselhaus2010}, which lifts the degeneracy of the $T_\pm$ and $T_0$ triplet states.
	
	Spin-orbit interaction creates various channels for phonon-assisted spin-flip processes. One class of such effects is driven by \textit{spin admixture} mechanisms, where the state with some nominal spin orientation gets a contribution of the opposite spin~\cite{Khaetskii2000a,Khaetskii2000b,pikus84,Mielnik-Pyszczorski2018a}. The other class contains direct spin-phonon coupling mechanisms~\cite{pikus84,Roth1960,Frenkel1991}, which in QDs are weaker compared to the channels due to the spin admixture~\cite{Khaetskii2000a,Khaetskii2000b,Mielnik-Pyszczorski2018a}.
	
	The phonon-induced transition rates between the states can be calculated using the Fermi golden rule
	\begin{align*}
		\Gamma(n \rightarrow m) &= \frac{2 \pi}{\hbar^2} \abs{n_{\mathrm{B}}(\omega_{nm}) + 1} \sum_{\kk,\lambda} \abs{G_{nm,\lambda}(\kk)}^2 \\
		&\quad \times [ \delta(\omega_{nm} - \omega_{k,\lambda}) + \delta(\omega_{nm} + \omega_{k,\lambda})],
	\end{align*}
	where $\omega_{nm} = (E_{n} - E_{m})/\hbar$ is related to the energy difference between the initial and final state, $n_{\mathrm{B}}(\omega)$ is the Bose-Einstein distribution, $\lambda$ denotes the phonon branch, and $\omega_{\kk,\lambda} = c_{\lambda} k$ with branch-dependent speed of sound $c_{\lambda}$. Finally, for the two-electron states 
	\begin{align*}
		G_{nm,\lambda}(\kk) =& \sum_{ii'jj'} c^{*(n)}_{ij} c^{(m)}_{i'j'} \Big[ F_{ii',\lambda}(\kk) \delta_{jj'} - F_{ij',\lambda}(\kk) \delta_{ji'}\\
		& - F_{ji',\lambda}(\kk) \delta_{ij'} + F_{jj',\lambda}(\kk) \delta_{ii'} \Big],
	\end{align*}
	where
	$$
		F_{ij,\lambda}(\kk) = \mel{\phi_i}{ H_{\mathrm{int}}(\kk,\lambda) e^{i \kk \rr}}{\phi_j},
	$$
	where $H_{\mathrm{int}}(\kk,\lambda)$ is the carrier-phonon interaction Hamiltonian for a single phonon mode.
	We take into account the deformation potential and piezoelectric electron-phonon couplings~\cite{Yu2010}. The detailed description of the carrier-phonon Hamiltonian is given in Ref.~\cite{Krzykowski2020}.

	\section{Results}
	\label{sec:results}	
	In this Section we discuss the calculated phonon-assisted relaxation rates. We also present the simulations of quantum kinetics for the two-electron singlet and triplet states.
	
	\subsection{Spin relaxation rates}
	\begin{figure}[tb]
		\begin{center}
			\includegraphics[width=85mm]{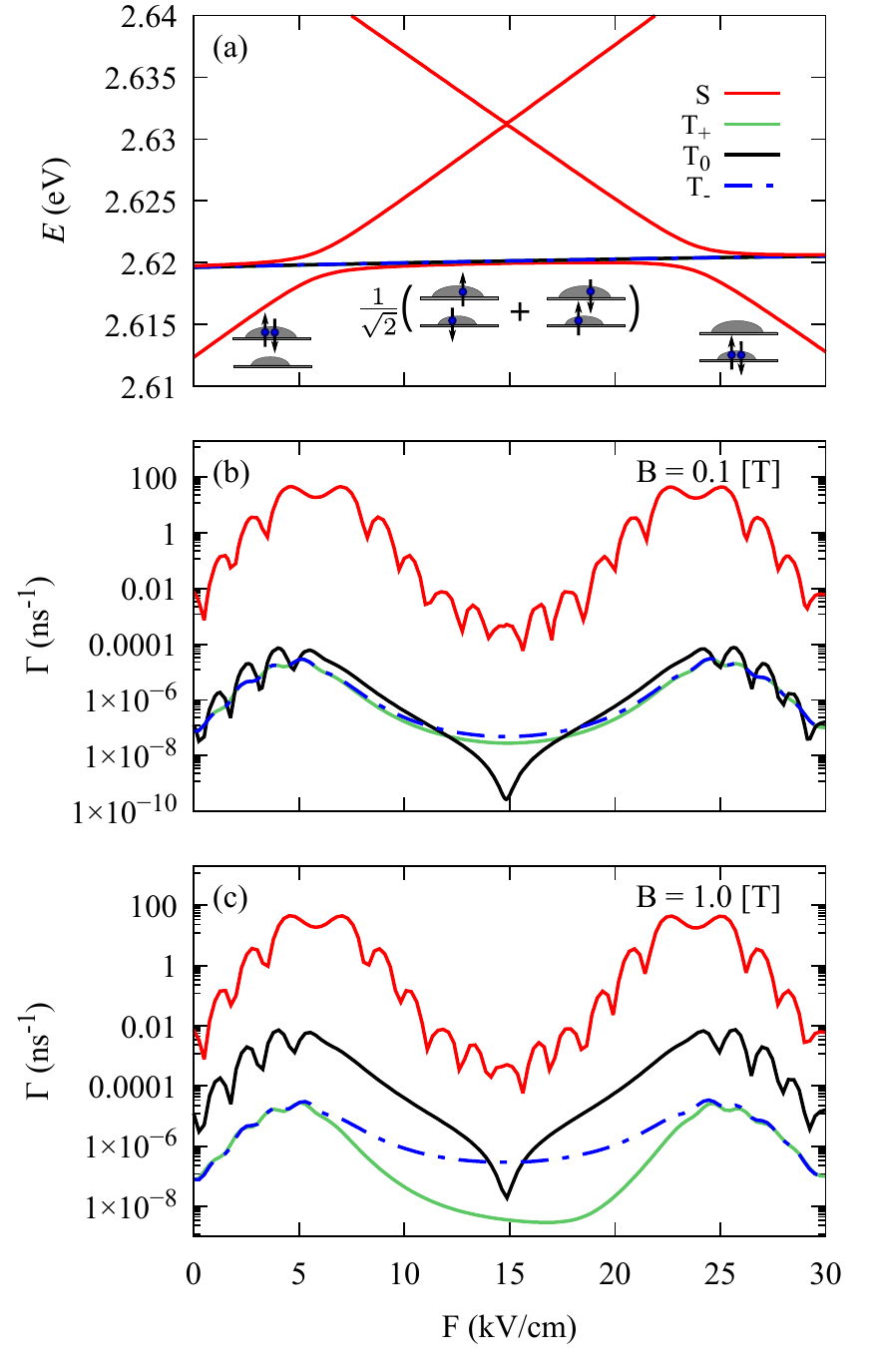}
		\end{center}
		\caption{\label{fig:energy}(Color online) (a) Two-electron energy levels as a function of electric field. The sketches present (approximate) spins projections and occupations in the lowest singlet state. (b,c) Phonon-assisted transition rates to the lowest singlet state from: the upper singlet (red solid line), triplet $T_+$ (green solid line), $T_0$ (black solid line), $T_-$ (blue dashed line). We assume $T=0$~K.
		}
	\end{figure}
	The energy spectrum for the lowest two-electron states is shown in Fig.~\ref{fig:energy}(a). At $F=0$, the ground state has (approximately) a configuration of $S(0,2)$, where both electrons are localized in the upper dot (which is assumed to be the larger one). The next three energy levels are nominally triplet states $T_\pm \equiv T_\pm(1,1)$, and $T_0 \equiv  T_0(1,1)$.  At $B=0$, the energy splitting between them is up to several neV, hence can be neglected. The next state forms nominally the $S(1,1)$ configuration. Finally, the energetically highest state is $S(2,0)$. 
	
	For the nonzero electric field, we obtained the well known structure of energy branches with three avoided crossings. 
	Since the positive electric field decreases (increases) the energy in the lower (upper) dot, the states can be tuned into resonance. In consequence, at $F \approx 5.81$~kV/cm and $F \approx 23.84$~kV/cm, there are pronounced avoided crossings, corresponding to the tunneling of a single electron between the dots. The anticrossing at $F \approx 14.85$~kV/cm is very narrow, which is due to two-particle character of the involved tunneling process.
	Furthermore, at $F \approx 14.86$~kV/cm there is a ``sweet spot", which corresponds to minimal difference (approximately $0.2$~meV) between the lowest singlet and triplet configurations.
	
	First, we calculate the phonon-assisted relaxation rates between the two lowest singlet states [red lines in Fig.\ref{fig:energy}(b,c)]. Depending on the magnitude of the electric field, the rates can describe different tunnel transitions: $S(1,1) \rightarrow S(0,2)$; $S(0,2) \rightarrow S(1,1)$; $S(2,0) \rightarrow S(1,1)$ or $S(1,1) \rightarrow S(2,0)$. The two pronounced maxima corresponding to the energy resonances from Fig.~\ref{fig:energy}(a) are related to the overlap between the wavefunctions (which is due to spatial delocalization). The oscillations of the relaxation rates result from the interference effects in the strongest confinement direction~\cite{bockelmann90}, and their period is related to the distance $D$ between the dots as $\propto 1/D$~\cite{wijesundara11,Gawarecki2010}. 
	Since the transitions between the singlet states are spin conserving, they do not significantly depend on magnetic field. The rates on the order of tens of ns$^{-1}$ for the dots separated by about $10$~nm are consistent with former predictions based on the single-band $\kp$ approximation~\cite{Gawarecki2010}. 
	
	Next, we calculate phonon-assisted transition rates from the triplet states to the lowest singlet state [Figs.~\ref{fig:energy}(b,c)]. 
	Depending on the electric field, such processes can either involve tunneling [$T(1,1) \rightarrow S(0,2)$ and $T(1,1) \rightarrow S(2,0)$], or conserve QD occupations [$T(1,1) \rightarrow S(1,1)$], or take an intermediate form. For the inter-dot tunnel transitions the rates oscillate, while no oscillations appear for the charge conserving processes.
	
	The transitions $T_{\pm} \rightarrow S(0,2)$ and $T_{\pm} \rightarrow S(2,0)$ can be considered as a single-electron tunneling accompanied by the spin-flip, while the other electron plays the role of a passive spectator. 
	Such transitions are driven (primarily) by the spin admixture mechanism related to the Dresselhaus spin-orbit coupling~\cite{Gawelczyk2019}. 	
	On the other hand, the rates related to the charge conserving processes $\Gamma(T_\pm \rightarrow S(1,1) )$ crucially depend on the difference between the spin admixtures of the states localized on different dots. 
	%
	%
	
	The relaxation processes $T_{0} \rightarrow S(0,2)$ and $T_{0} \rightarrow S(2,0)$ exhibit different behavior compared to those involving the $T_{\pm}$. We have verified that, in this case, the spin admixture mechanisms play minor role. Instead, the dominant transition channel is related to the difference of the single-particle $g$-factors for the states localized on different dots. This is a real-space spin-orbital effect connecting the position to the spin degree of freedom, which was shown to be an important factor determining transport properties of a double quantum dot\cite{Mutter2020}.  It can be interpreted with the effective Zeeman Hamiltonian
	$$
	H_{\mathrm{Z}} = \frac{1}{2} \mu_B g_1 \sigma^{(1)}_z B_z + \frac{1}{2} \mu_B g_2 \sigma^{(2)}_z B_z,
	$$
	where $g_{1/2}$ and $\sigma^{(1/2)}_z$ are $g$-factors and the ($z$-th) Pauli matrices defined for a given single-particle orbital ($1$ or $2$). While such a Hamiltonian conserves the projection of the total angular momentum $J_z$, it does not commute with the $J^2$ operator, hence mixes the states belonging to different representations of the rotation group. Such an effect takes place only for many-body configurations. In the case of two electrons, the $T_\pm$ states are unaffected because $\mel{S}{H_Z}{T_\pm}=0$. On the other hand, $\mel{S(1,1)}{H_Z}{T_0}=\frac{1}{2} \mu_B (g_1 - g_2) B_z$ which mixes the states. This rise a spin-admixture that leads to a phonon-assisted relaxation~\cite{Khaetskii2000a,Khaetskii2000b,Mielnik-Pyszczorski2018a}. For the QDM system considered here,  $g_{\mathrm{u}} \approx -0.89$ vs. $g_{\mathrm{l}} \approx -0.71$ for the upper and the lower dot respectively, opening a significant $T_0 \rightarrow S$  relaxation channel. 
    In consequence, the transition rate $\Gamma(T_{0} \rightarrow S(1,1))$ shows a minimum at $F\approx 14.7$~kV/cm, which is near the point where the single-particle g-factors are equal [$g \approx (g_{\mathrm{u}} + g_{\mathrm{l}})/2$ ] due to the delocalization of the electron states. This is very close to, although not exactly coinciding with, the S-T ``sweet spot'' at $F \approx 14.86$~kV/cm.
	
	\begin{figure}[tb]
		\begin{center}
			\includegraphics[width=85mm]{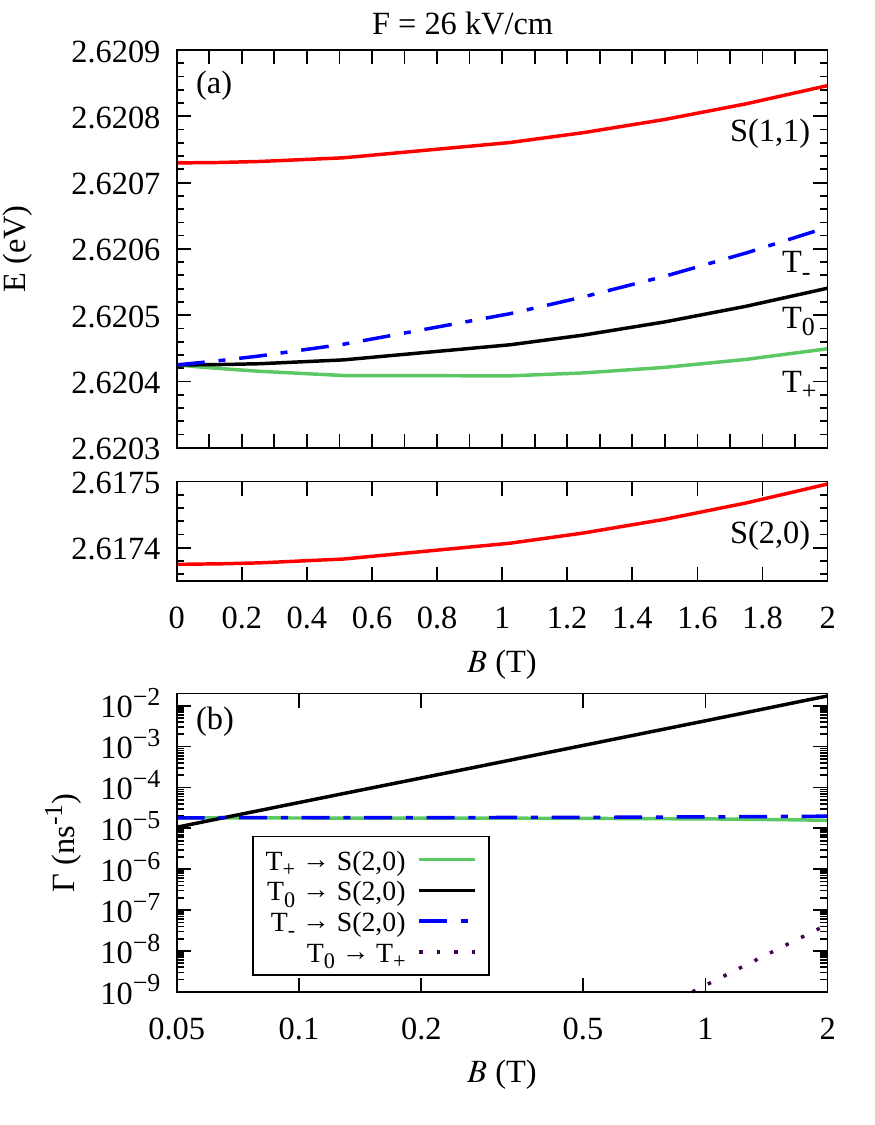}
		\end{center}
		\caption{\label{fig:Bfield}(Color online) Energy branches (a) and phonon-assisted relaxation rates at $T=0$~K (b) as a function of axial magnetic field $B$, at $F=26$~kV/cm. 
		}
	\end{figure}	
	The magnetic-field dependence of the lowest two-electron states at $F=26$~kV/cm is shown in Fig.~\ref{fig:Bfield}(a). The energies of all the states contain a diamagnetic contribution $\propto B^2$. In addition, the energies of the spin-polarized states $T_\pm$ show a Zeeman shift. Since the single-electron g-factors are negative, the $T_{+}$ state has a lower energy than $T_{-}$.
	The calculated phonon-assisted relaxation rates between the states are shown in Fig.~\ref{fig:Bfield}(b). 
	For the considered magnetic-field range, one can approximate the power dependence $\Gamma(T_{\pm} \rightarrow S(2,0)) \approx \mp a B + c$. The dominant component $c$ depends (mainly) on the Dresselhaus spin-admixture mechanism~\cite{Gawelczyk2019}.
	The relaxation from the $T_{0}$ state follows $ \Gamma(T_{0} \rightarrow S(2,0)) \approx a_0 B^2$, where the parameter $a_0$ mainly depends on the difference between the single-particle g-factors of the Zeeman doublets $a_0 \propto (g_\mathrm{u}-g_\mathrm{l})^2$, consistent with the discussion in terms of the effective Zeeman Hamiltonian, presented above. Finally, the relaxation $T_0 \rightarrow T_+$ can be interpreted as a single-electron spin-flip in the individual dots. The rate $\Gamma(T_0 \rightarrow T_+)$ depends on magnetic field as $\propto B^5$, which is consistent with the results for one electron in a single QD~\cite{Khaetskii2000a,Khaetskii2000b,Mielnik-Pyszczorski2018a}. 
		
	\begin{figure}[tb]
		\begin{center}
			\includegraphics[width=85mm]{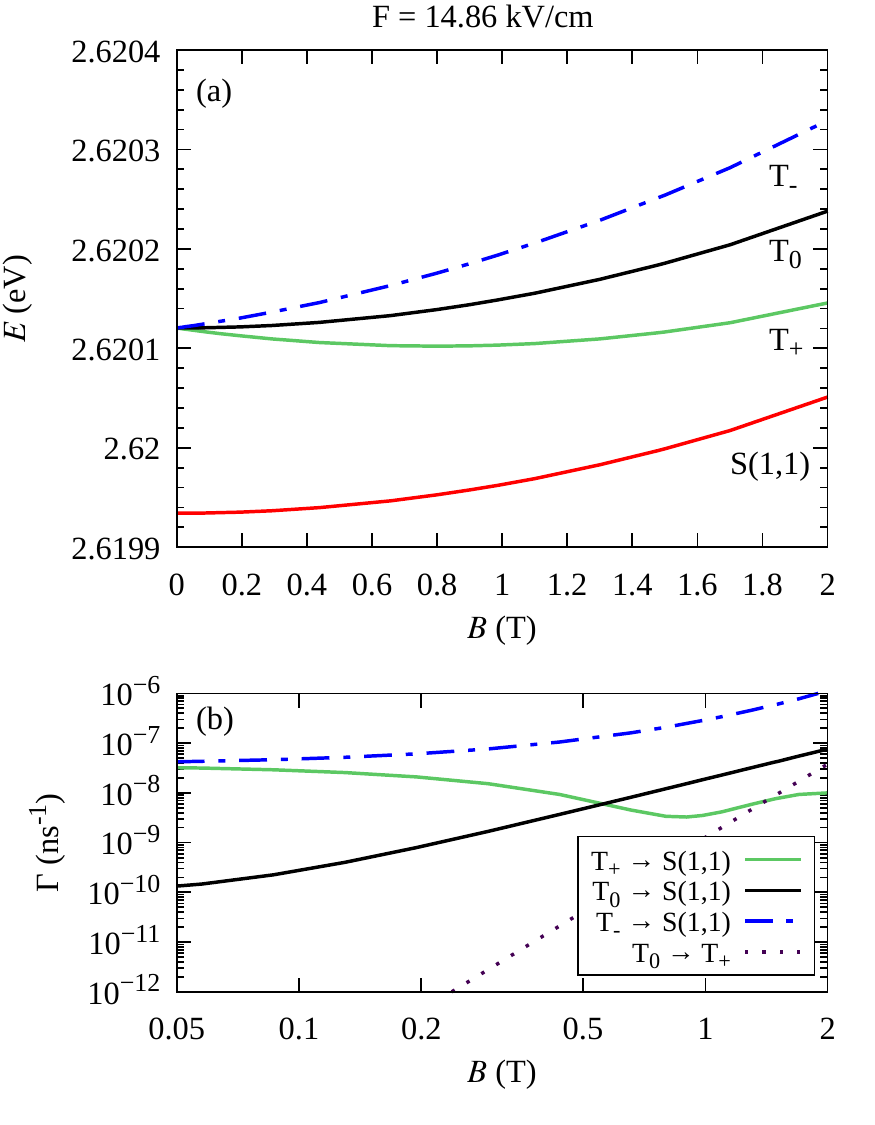}
		\end{center}
		\caption{\label{fig:Bfield_15}(Color online) Energy branches (a) and phonon-assisted relaxation rates at $T=0$~K (b) as a function of axial magnetic field $B$, at the ``sweet spot'' $F=14.86$~kV/cm. 
		}
	\end{figure}	
	Next, we calculate the energy branches [Fig.~\ref{fig:Bfield_15}(a)] and the relaxation rates [Fig.~\ref{fig:Bfield_15}(b)] for the electric field corresponding to the sweet spot ($F=14.86$~kV/cm).
	The observed dependencies of the rates result from the interplay of various SO mechanisms (which can interfere constructively or destructively), and Zeeman energy splitting (important for the considered energy scale). 
	The triplet-singlet transitions are several orders of magnitude slower compared to the rates shown in Fig.~\ref{fig:Bfield}(b).
	This is partly related to the fact, that the phonon spectral density decreases at low frequencies, and the energy differences between the S and T states are now relatively small.   
	In addition, a strong delocalization of the electron states decreases some of the spin-orbit coupling mechanisms, giving rise to $T_\pm \rightarrow S(1,1)$ and $T_0 \rightarrow S(1,1)$ relaxation.
	On the other hand, the magnitude of $\Gamma(T_0 \rightarrow T_+)$ is very similar (with the difference of about $10$\%) to the case of $F=26$ ~kV/cm. This is due to the fact, that the $T_0 \rightarrow T_+$ transition involves spin-flips in the individual dots, which are not sensitive to the electric field.

	\subsection{Relaxation kinetics}
	\begin{figure}[tb]
		\begin{center}
			\includegraphics[width=90mm]{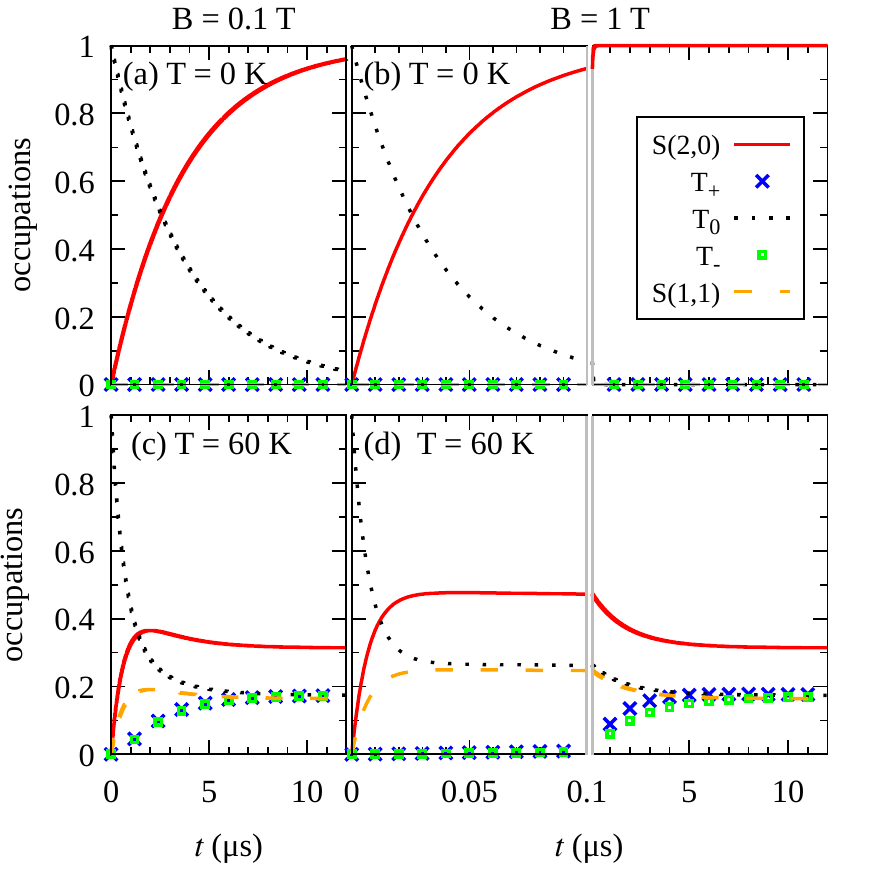}
		\end{center}
		\caption{\label{fig:time}(Color online) Time dependencies of the occupations at $F=26$~kV/cm.
		}
	\end{figure}	
	To investigate the kinetics of the system, we numerically solve the Master equation assuming the initial occupation of the $T_{0}$ state. We focus on the regime of the tunnel transitions, where we took $F=26$~kV/cm, and we consider magnetic fields of $B = 0.1$~T and $B = 1$~T.  At $T=0$~K [Fig.~\ref{fig:time}(a,b)] the evolution is exponential, where the only significant process is the direct relaxation $T_{0} \rightarrow S(2,0)$. However, for a non-zero temperature, the picture becomes more complicated. The transitions to upper states become possible, and the final occupations form a temperature-dependent equilibrium (according to the Gibbs distribution). Furthermore, the Bose-Einstein distribution of phonons enhances the rates between states separated by a small energy difference. As shown in Figs.~\ref{fig:time}(c,d), for $T=60$~K the evolution of occupations is no longer exponential and involves more states. The relaxation $T_{0} \rightarrow S(2,0)$ to the lowest singlet state can occur directly, but also through  $T_{0} \rightarrow  S(1,1) \rightarrow  S(2,0)$.  Although the phonon-assisted transitions $T_{0} \leftrightarrow T_{\pm}$ are negligible, the states  $T_{\pm}$ get occupied through $T_{0} \rightarrow  S(1,1) \rightarrow  T_{\pm}$, and $T_{0} \rightarrow  S(2,0) \rightarrow  T_{\pm}$ transitions. For a weak magnetic field [Fig.~\ref{fig:time}(c)], the occupation dynamics of all the triplet states exhibit comparable timescales. On the other hand, the higher magnetic field [Fig.~\ref{fig:time}(d)] leads to two distinct regimes: the fast transitions $T_{0} \rightarrow S(2,0)$ and $T_{0} \rightarrow S(1,1)$ in the first stage of the evolution, then slow transitions to the $T_\pm$. Such behavior results from different magnetic-field dependencies of the involved transition rates [see Fig.\ref{fig:Bfield}(b)].
	
	\begin{figure}[tb]
		\begin{center}
			\includegraphics[width=90mm]{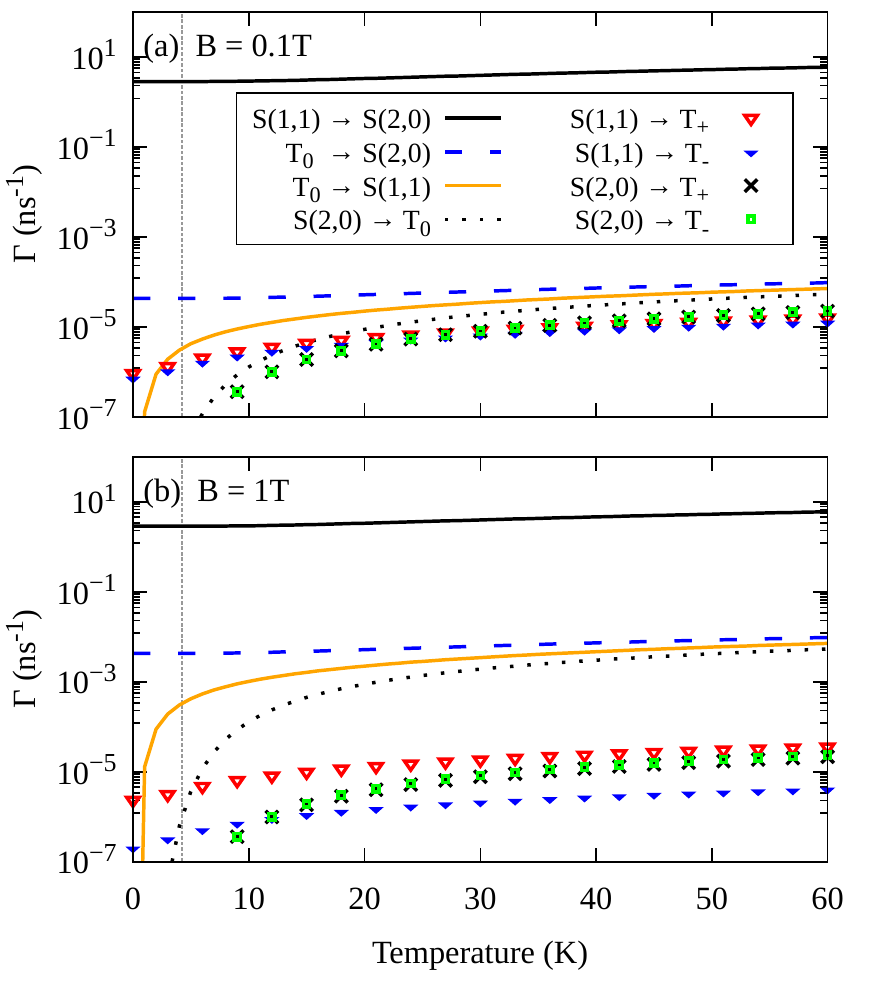}
		\end{center}
		\caption{\label{fig:temp}(Color online) Temperature dependence of phonon-assisted relaxation rates at $F=26$~kV/cm. The vertical lines denote $T = 4.2$~K.
		}
	\end{figure}	
	In order to quantitatively assess the importance of distinct transition channels for the kinetics presented in Fig.~\ref{fig:time}, we calculate the relaxation rates as a function of temperature (Fig.~\ref{fig:temp}). For low temperatures, the relaxation is clearly dominated by the direct transition $T_{0} \rightarrow S(2,0)$. However, with increasing temperature and at low magnetic fields, the transitions involving the $T_\pm$ start to play important role in the system dynamics.
	On the other hand, at higher magnetic fields [$B=1$~T in Fig.~\ref{fig:temp}(b)], the transitions from/to the $T_{\pm}$ states are very slow compared to the $T_0 \rightarrow S(2,0)$ in the whole range of temperatures.

	\section{Conclusions}
	\label{sec:concl}
	We have studied quantum kinetics of two electrons in a quantum dot molecule. With a realistic model of the QD system geometry, $8$-band $\kp$ method, and configuration-interaction approach, we have calculated two-electron states. We have investigated the phonon-assisted transitions between the triplet and singlet states in the presence of external magnetic and electric fields. We have considered the triplet-singlet transitons accompanied by tunneling as well as the case of occupation conserving relaxation. We have identified channels of the triplet-singlet relaxation that become important in different parameter regimes. While for weak magnetic fields the tunnel transitions related to spin-admixture mechanisms are dominating, the regime of moderate and strong magnetic fields favors another mechanism related to the difference between the electron g-factors in the dots. We have also shown, that near the ``sweet spot'', lifting of this mechanism leads to a considerably longer lifetime of the triplet $T_0$ state. Finally, we have demonstrated a non-exponential quantum kinetics resulting from the
	interplay of various direct and sequential processes contributing to the relaxation.

	\acknowledgments
	This work was supported from the Polish National Science Centre (NCN) under Grant
	No. 2016/23/G/ST3/04324.
	Part of the calculations have been carried out using resources provided by Wroc{\l}aw Centre for Networking and Supercomputing (\url{http://wcss.pl}), Grant No.~203.
	We are also grateful to Micha{\l} Gawe{\l}czyk for sharing his implementation of the blur algorithm.
	

\bibliographystyle{prsty}
\bibliography{abbr,../library.bib}
	
\end{document}